# Superfluorescent scintillation from coupled perovskite quantum dots

Shaul Katznelson[1†], Shai Levy[1,2†], Alexey Gorlach[1†], Nathan Regev[1†], Michael Birk[1], Chen Mechel[1], Offek Tziperman[1], Roman Schuetz[1], Rotem Strassberg[1,2], Georgy Dosovitsky[1], Charles Roques-Carmes[3], Yehonadav Bekenstein[1,2], and Ido Kaminer[1]

[1]*Solid State Institute, Technion – Israel Institute of Technology, Haifa 3200003, Israel*
[2]*Faculty of Materials Science and Engineering, Technion – Israel Institute of Technology, Haifa 3200003, Israel*
[3]*E. L. Ginzton Laboratory, Stanford University, Stanford, California 94305, US*
† *equal contributors*
kaminer@technion.ac.il

**Scintillation, the process of converting high-energy radiation to detectable visible light, is pivotal in advanced technologies spanning from medical diagnostics to fundamental scientific research. Despite significant advancements toward faster and more efficient scintillators, there remains a fundamental limit arising from the intrinsic properties of scintillating materials. The scintillation process culminates in spontaneous emission of visible light, which is restricted in rate by the oscillator strength of individual emission centers. Here, we observe a novel collective emission phenomenon under X-ray excitation, breaking this limit and accelerating the emission. Our observation reveals that strong interactions between simultaneously excited coupled perovskite quantum dots can create collective radioluminescence. This effect is characterized by a spectral shift and an enhanced rate of emission, with an average lifetime of 230 ps, 14 times faster than their room temperature spontaneous emission. It has been established that such quantum dots exhibit superfluorescence under UV excitation. However, X-ray superfluorescence is inherently different, as each high-energy photon creates multiple synchronized excitation events, triggered by a photoelectron and resulting in even faster emission rates, a larger spectral shift, and a broader spectrum. This observation is consistent with a quantum-optical analysis explaining both the UV-driven and X-ray-driven effects. We use a Hanbury-Brown-Twiss $g^{(2)}(\tau)$ setup to analyze the temperature-dependent temporal response of these scintillators. Collective radioluminescence breaks the limit of scintillation lifetime based on spontaneous emission and could dramatically improve time-of-flight detector performance, introducing quantum enhancements to scintillation science.**

## Introduction

Detecting high-energy photons, such as X-rays and gamma rays, is a common challenge in science and technology. Many predominant detection methods rely on scintillation – a process in which an energetic quantum of ionizing radiation produces multiple lower-energy photons, primarily in the optical regime, that can be readily detected by various photodetectors[1]. Scintillation has a wide range of applications. In high-energy physics, scintillators were crucial elements in detectors that enabled the discovery of the Higgs boson at LHC, CERN[2,3]. In medical imaging, they enable a vast range of diagnostic tools such as positron emission tomography (PET) and X-ray computed tomography (CT)[4–6]. Other applications of scintillators include homeland security[7], non-destructive testing and materials characterization[8,9], nuclear industry[10–13], and oil well logging[14]. Although different applications have specific requirements for scintillators' characteristics, they all rely on the same underlying process, and typically require the scintillation light to be as bright and fast as possible[15–17].

One important example of where lifetime improvements are especially crucial is for higher-resolution imaging in time-of-flight measurements. Famously in positron emission tomography, shortening the scintillation lifetime could greatly improve imaging capabilities, enabling the detection of smaller tumors at earlier stages and earlier diagnosis of neurodegenerative diseases[18,19]. For this goal and many others, ongoing research efforts strive to improve scintillation yield and lifetime[16,17,20].

New strategies for improving scintillator performance[21] involve efforts across areas of materials science, chemistry, and physics. These efforts include macroscopic structuring of scintillators[1], nanophotonic scintillators[22–29], and quantum-dot (QD) scintillators[30,31,31–42]. While each of these concepts holds great promise, they all still rely on the intrinsic properties of the scintillator compounds they are made of – such that the oscillator strength of *individual* emission centers limits their performance. Achieving scintillation *collective* emission could bypass these constraints and enable previously inaccessible regimes of ultrafast scintillation.

The process of collective emission is based on a spontaneous build-up of correlated emission among closely spaced emitters. The first proposal of correlated spontaneous emission was Dicke's superradiance[43]. This phenomenon can be extended to describe superradiant emission by free electrons[44–46] and used as a mechanism for quantum light generation[47]. Collective emission can emerge spontaneously from "incoherent" excitations at shorter wavelengths – termed superfluorescence[48] – as observed in atomic and molecular vapors[49,50], solid-state films[51–53], quantum dots (QDs)[54,55], and organic molecular aggregates[56–60].

A powerful new platform for studying superfluorescence is halide perovskite QDs[54], which are renowned for their optoelectronic properties[61–66]. Since initial observation of collective emission in CsPbBr$_3$ QD superlattices[54], research has proliferated, with reports of various collective emission phenomena from different CsPbBr$_3$ QDs[67–70], and also from thin film halide perovskites[52,53]. This emerging field now includes reports such as room-temperature superfluorescence[53], single-photon superradiance[69], dipole-dipole-interaction-mediated superfluorescence[70], and an order-disorder phase-transition[71]. So far, all these reports of collective emission were under optical (visible and UV) excitations.

Following on our earlier report of shortened lifetime under X-ray scintillation from QD superlattices[72], we now complete the picture, showing this effect to be a form of collective radioluminescence. Recently we learned of a related effect reported in isolated QDs[73]

Here, we observe X-ray-driven collective emission, which we measure from superlattices of perovskite QDs, reaching average radioluminescence lifetimes of 230 ps at 80 K. Collective radioluminescence manifests itself as a separate, red-shifted spectral peak, similarly to previously observed UV-driven superfluorescence, yet with a larger red-shift and a broader spectrum. We report that QD superlattices display these fast superfluorescence features when driven by either blue light (up to 425 nm), UV (375 nm), UVC (222 nm), or X-rays (8 keV).

We experimentally study the temporal and spectral behavior of this collective scintillation using temperature-dependent Hanbury-Brown-Twiss (HBT) correlation measurements[74] and emission spectra, for various samples and excitation conditions. Emission lifetime no longer depends only on the individual emitter characteristics but also on interactions between these emitters, providing an avenue for much shorter scintillation lifetimes.

By comparing the spectral and temporal features under UV and X-ray excitation, we identify that X-ray excitation results in an increased spectral splitting, spectral width, and emission rate compared to the conventional blue light or UV excitation. To better understand these differences between UV and X-ray excitations, we develop a two-step theory that describes (1) the excitation of perovskite QDs using Monte-Carlo simulation[75–77] of energetic photoelectrons inside the material (Fig. 1b) and then (2) describes the consequent collective emission process of optical photons by exploiting quantum-optical theory (Fig. 1a,c), based on Lindblad master equation[78]. The same quantum optical theory explains the features of collective emission under both UV and X-ray excitations. The explanation for the differences between collective emission driven by UV and by X-ray is the much higher density of simultaneously excited QDs in the latter.

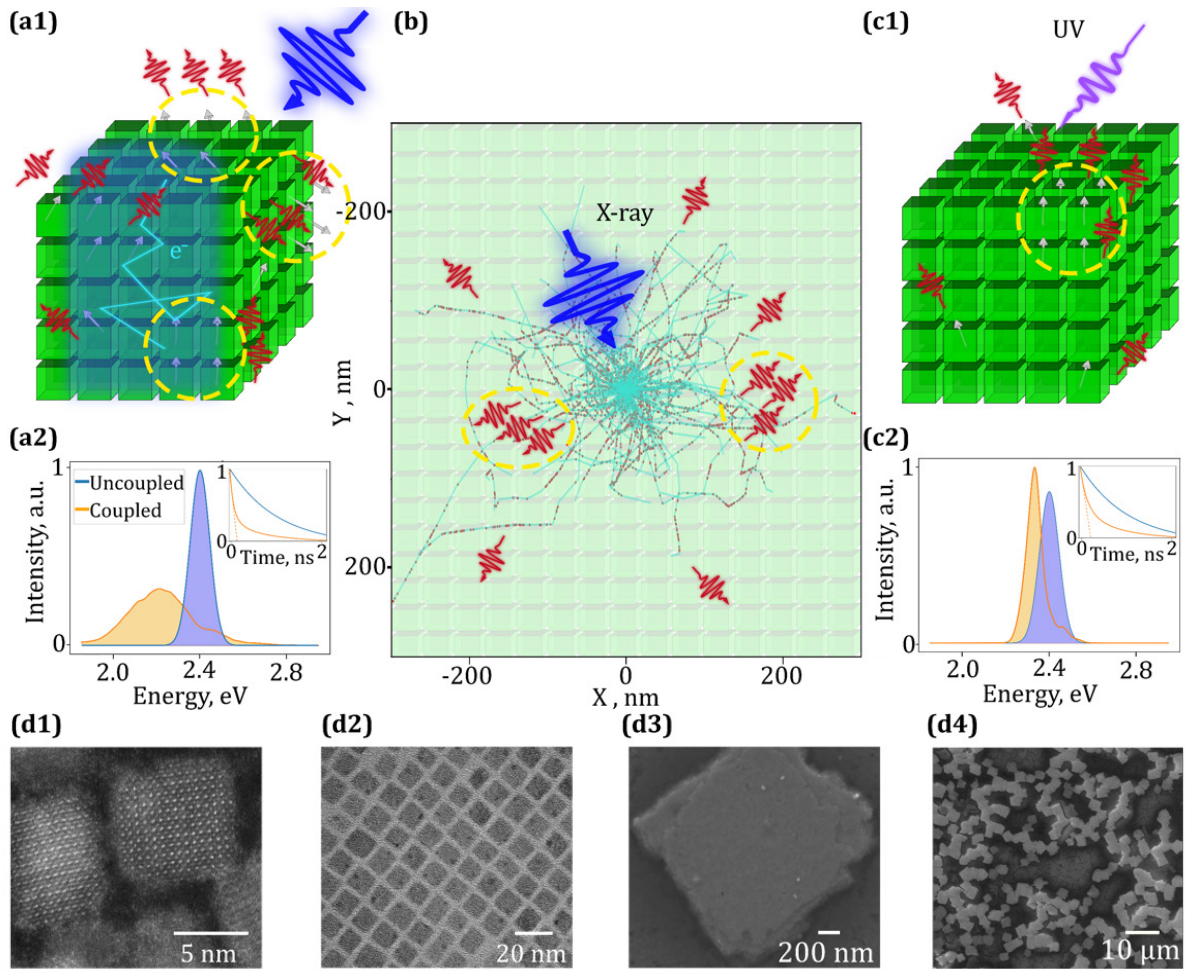

**Figure 1: Theory and simulations of collective emission from CsPbBr$_3$ QD superlattices under X-ray and UV excitations. (a1)** Diagram: Each X-ray photon generates an energetic electron that creates multiple collective excitations inside the superlattice. **(a2)** Theoretically calculated emission spectrum and lifetime under X-ray excitation, comparing uncoupled (blue) and coupled (orange) emission. **(b)** Monte-Carlo simulation of energetic electrons (cyan lines) generated by X-ray excitation, demonstrating multiple excitations of neighboring QDs (red dots), which explains the broader spectrum and shorter lifetime of X-ray collective emission. **(c1)** Diagram: Each UV photon generates a single collective excitation that leads to the conventional superfluorescence. **(c2)** Theoretically calculated spectrum and lifetime under UV excitation, comparing uncoupled (blue) and coupled (orange) emission. Both types of emission, under both UV and X-rays, were simulated using the Lindblad master equation, see supplementary information (SI) section 8. The QDs transition energies were modeled by a Gaussian distribution of $\sigma_E = 100$ meV around $\Delta E = 2.45$ eV. The dipole-dipole

interaction strengths between nearest QDs were modeled as constant with $J_0 = 100$ meV under UV excitation. The X-ray excitation creates simultaneous neighboring excitations modeled by higher variability in the interaction strengths, with Gaussian distribution of $\sigma_J = 100$ meV around $J_0 = 150$ meV. **(d1)** HAADF-STEM micrograph of a single CsPbBr$_3$ QD with atomic resolution. **(d2)** TEM micrograph of QDs ordered in a superlattice. **(d3)** SEM micrograph of a single superlattice, containing multiple QDs. **(d4)** SEM micrograph of a large area containing multiple cubic superlattices.

**Results**

We start by describing the theory predicting the expected optical behavior of a superlattice of CsPbBr$_3$ QDs under excitation of X-ray or UV photons. We first present the universal features expected to appear in both energy regimes, and then explain the differences that make the X-ray-driven emission stand out. In both regimes, we model the emission process using Lindblad master equation of two-level systems coupled via dipole-dipole interactions – see supplementary information (SI) section 8 – yielding spectra and emission rates as shown in Fig. 1(a2,c2). For both X-ray and UV excitations, we assume the same distribution of QD transition energies. This distribution is responsible for the inhomogeneous broadening of the uncoupled emission, observed in room temperature, which is similar under both X-ray and UV excitations (Fig. 1a,c blue curves).

For low temperatures, the dipole-dipole interactions between neighboring QDs synchronize them, such that each excitation can be shared among several QDs. This synchronization leads to collective emission (Fig. 1a2,c2 orange curve), characterized by the shorter lifetime and the red-shifted spectral peak[56,60]. This prediction is consistent with previous experiments reporting superfluorescence[54,67–69,79] and with our measurements here.

These important differences between emission driven by UV and X-rays appear at low temperatures, since the collective emission also depends on the density of excitations. A Monte-Carlo simulation[75–77,80] reveals the different density of excitations between the two energy regimes: Each UV photon can only create a single excitation, as the 8nm CsPbBr$_3$ QDs (Fig. 1d) have an optical band gap around 2.45 eV[61], whereas an X-ray photon creates multiple excitations in a cascade process (Fig. 1b). In the scintillation cascade process, the X-ray photon is first absorbed by the material, generating an energetic photoelectron. This electron then moves and excites many QDs in its trajectory, even up to hundreds depending on its energy, and the rate of energy deposition into a material increases to the end of the track, as the electron decelerates[18,81]. In the case of 8 keV X-ray photons, the Monte Carlo simulation predicts the mean distance between excitations to be 50 nm, creating around each excited QD an average of 2 excited neighbor QDs (see SI section 9). In contrast, for our UV excitation, the mean distance between excited QDs is far larger, making it extremely unlikely to have any two neighboring QDs simultaneously excited.

This higher density of excitations does not affect the emission at room temperature, due to the weak dipole-dipole interactions between neighboring QDs, causing the uncoupled emission under X-rays to be similar to that excited by UV (Fig. 1a2,c2 blue curves). In contrast, at lower temperatures, the stronger dipole-dipole interactions cause the multiple neighboring excitations to affect each other. We model this neighbor-coupling as an effective increase of the dipole-dipole interaction strengths and variability (see SI section 8). The Lindblad master equation theory then predicts faster emission rates, a larger red-shift, and a broader spectrum for X-ray-driven collective emission relative to the same effect under UV excitation (Fig. 1a2, c2 yellow curves). These predictions are consistent with our X-ray measurements and with comparison to UV measurements on the same samples.

To experimentally test this behavior, we synthesized 8 nm monodispersed colloidal $CsPbBr_3$ QDs, as required for the formation of highly ordered assemblies[82], using the procedure from Ref. [83]. Drop casting the deposited QDs self-assembled on the substrate creates three-dimensional cubic-packed rectangular superlattices (Fig. 1d1-4, and SI section 1).

To probe the collective emission phenomenon, we employ micro photoluminescence and X-ray radioluminescence systems, at a range of temperatures down to 80 K. The radioluminescence measurement use an 8 keV X-ray tube attached to a polycapillary lens with a spot size of 50 μm. The temperature of the samples can be varied between $80 - 300$ K using a cryostat with windows enabling transmission experiments along an optical axis. The optical photons emitted from the sample are collected by an objective lens in which an iris allowing us to change the spot size. The light is coupled to an optical fiber, transporting it to either spectrum or lifetime measurements (see SI section 2).

We measured the luminescence of the QDs superlattices under several excitation energies (see SI section 2): Xenon lamp coupled with a notch filter (325 − 425 nm), pulsed blue laser (405 nm), pulsed UV laser (375 nm), UVC KrCl excimer lamp (222nm), and X-ray tube (8 keV). For the UV and X-ray excitations, we scanned over a temperature range 80 − 300 K and observed in both cases the gradual emergence of coupled emission, characterized by a red-shifted spectral peak with enhanced emission rate.

Figure 2 presents the measured spectra: showing two distinct spectra peaks, identified as spontaneous emission from individual uncoupled QDs, centered at 2.45 eV, and collective emission from coupled QDs, red-shifted by 320 meV under X-ray excitation (Fig. 2a) and 60 meV under UV excitation (Fig. 2c). The relative weight of collective emission is 77% in Fig. 2a and 51% in Fig. 2c. Other samples show typical red-shifts of 80-200 meV and typical yields of 20-60% under X-ray excitation. Collective emission typically appears below 180 K (Fig. 2b,d) and even 220 K in certain samples (see SI sections 3 and 4).

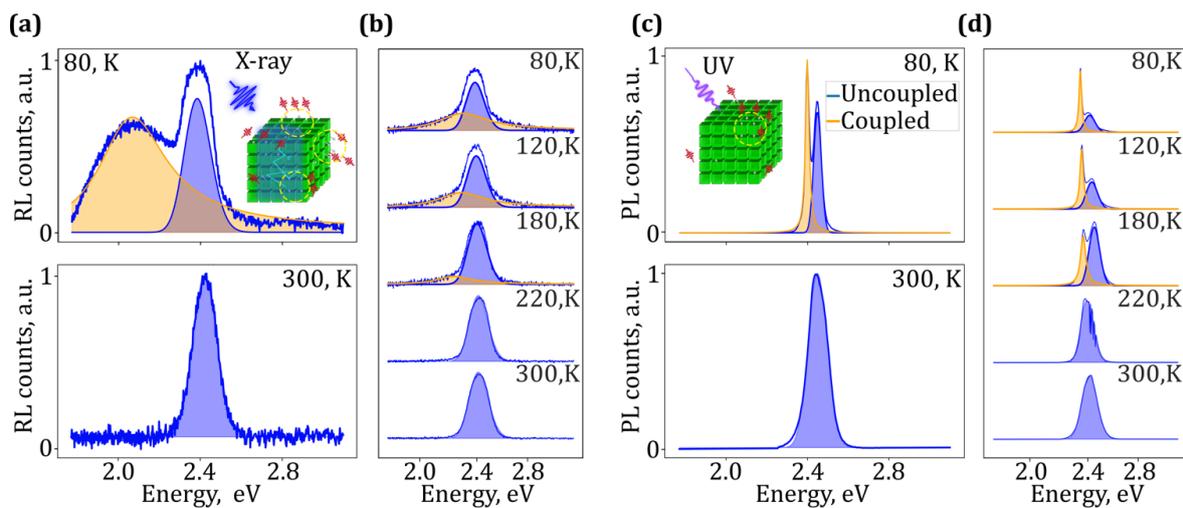

**Figure 2: Spectra of collective radioluminescence (RL) and photoluminescence (PL). (a)** Emission spectra for the QD superlattices at 80 K (top) and 300 K (bottom) under X-ray excitation. At low temperatures, a broad red-shifted peak (yellow) appears next to the spontaneous emission peak (light blue) that exists also at room-temperature. Compared to the UV excitation, the red-shifted peak is broader and its shift is larger (both reaching up to a few hundreds of meV, rather than tens of meV). **(b)** Spectra at a range of temperatures under X-ray excitation for a different sample, emphasizing the variability of the effect, here showing a broader spectral peak with a smaller red-shift. **(c)** Emission spectra as in (a), under UV laser excitation. **(d)** Spectra as in (b), under UV laser excitation, emphasizing the variability of the effect, not so much in the width of the red-shifted peak, but more in its relative weight and red-shift. Similar spectral features are measured for different substrates; examples here are on Kapton-Au for (a,b,d) and Kapton for (c).

All measurements on all samples show the spectral peak of collective radioluminescence to be broader than that of collective photoluminescence, with a larger red-shift. This difference is expected from theory, as each UV photon can produce at most a single excitation while each X-ray photon can produce multiple simultaneous excitations. The increased spectral broadening and red-shift arise from the coupling among these multiple excitations via the strong dipole-dipole interactions of neighboring QDs. The substantial spectral broadening arises from the greater variability in coupling strengths among multiple excitations, which can be further amplified by defects such as domain boundaries, residual strain from assembly, and an angular disorder in QD alignment[71,84–87].

In addition to the spectral changes, the collective emission emerging from coupled QDs shows a substantial lifetime reduction compared to spontaneous emission from uncoupled QDs. Fig. 3a,b display the lifetime obtained from second-degree photon correlations $g^{(2)}(\tau)$, measured using Hanbury-Brown-Twiss interferometry[74,88]. In Fig. 3a, we extract the "fast" and "slow" emission lifetimes, $0.24 \pm 0.02$ ns and $0.74 \pm 0.12$ ns, corresponding to the coupled and uncoupled spectral components of Fig. 2a, respectively. Evidently, the relative weights of the two exponents in the fit of Fig. 3a are similar to the relative weights of the uncoupled and coupled spectral peaks of Fig. 2b at 80 K of 56% coupled emission and 44% uncoupled emission. This correspondence is consistent across all our measurements at different temperatures, collection areas, and substrates (see SI sections S, 4, and 6). Compared to the $3.35 \pm 0.07$ ns emission lifetime at 300 K (Fig. 3b), the coupled emission at 80 K is 14 times faster.

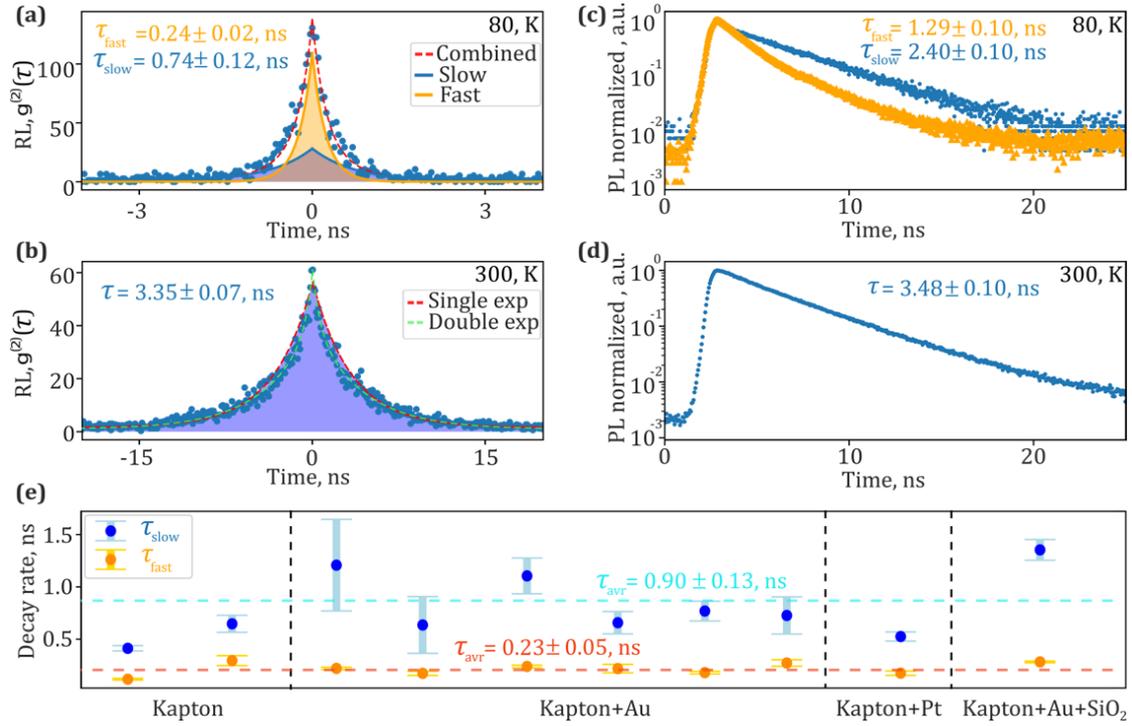

**Figure 3: Lifetime of collective radioluminescence (RL) and photoluminescence (PL). (a)** $g^{(2)}(\tau)$ curve at 80 K under X-ray excitation. Two lifetimes are extracted: fast ($\tau_{fast} = 0.24 \pm 0.02$ ns, orange) and slow ($\tau_{slow} = 0.74 \pm 0.12$ ns, blue). **(b)** $g^{(2)}(\tau)$ curve for emission from the same spot as in (a) at 300 K under X-ray excitation, yielding a single lifetime ($\tau = 3.35 \pm 0.07$ ns). We also denote (dashed curve) the slightly better fit to two lifetimes, with the faster component representing 6% of the emission with lifetime of ~0.5 ns, providing a weak indication for a small component of collective emission, which will be investigated in future work. **(c)** Standard emission curves at 80 K under pulsed UV laser excitation, showing two lifetimes: fast ($\tau_{fast} = 1.29 \pm 0.01$ ns, orange) corresponding to the narrow redshifted peak in Figure 2(c), and slow ($\tau_{slow} = 2.40 \pm 0.01$ ns, blue) corresponding to the broader peak. **(d)** Standard emission curve at 300 K under pulsed UV laser excitation, showing a single slow lifetime ($\tau = 3.48 \pm 0.01$ ns). **(e)** Summary of lifetimes measured at 80 K under X-ray excitation for 10 measured samples on different substrates. The fast ($\tau_{fast} = 0.23 \pm 0.05$ ns) and slow ($\tau_{slow} = 0.90 \pm 0.13$ ns) lifetimes vary across samples, yet consistently demonstrate an approximately 4-fold enhancement of the coupled emission rate relative to uncoupled emission at the same temperature. The substrates are Kapton-Au for (a,b) and Silicon for (c,d).

For comparison, Figs. 3c,d display the time-resolved photoluminescence (TRPL) measurements under UV laser excitation, showing reduced lifetime at 80 K relative to 300 K, due to collective emission, as expected from previous works[54]. The lifetimes of the uncoupled emission (2.45 eV) and coupled emission (2.39 eV) are $2.40 \pm 0.01$ ns and $1.29 \pm 0.01$ ns, measured using a single-exponent fitting for the filtered emission at the spectral peaks in Fig. 3c. Interestingly, collective emission excited by X-rays is faster than that excited by UV, as expected by theory (insets in Fig. 1a2,c2), due to the stronger interactions induced by multiple excitations produced by each X-ray photon.

**Discussion**

Looking at the bigger picture, various mechanisms of collective emission have now been observed for perovskite quantum dots (e.g. superfluorescence, single-photon superradiance, and giant oscillators strength[67–70]). Efforts are still ongoing to fully understand the intrinsic mechanism[89]. Our work contributes to this broader discussion by extending (by 3-4 orders of magnitude) the energy regimes under which signature features of collective emission have been measured. The fact that the same theoretical framework captures the effects under both UV and X-ray excitations broadens the scope and improves our understanding of these phenomena.

We proceed to examine other emission mechanisms prevalent in QDs and contrast their effects to those of collective emission. The (average) values of the observed red-shifts, of ~85 meV under X-ray excitation and ~50 meV under UV excitation, indicate strong dipole-dipole interaction between adjacent QDs in the superlattice. For contrast, typical spectral shifts due to biexcitons, trions, or Auger recombination are up to a few tens of meV and often much smaller[54,90–92]. It is possible that the simultaneous multi-excitations created by each X-ray photon combines superfluorescence with such multi-exciton phenomena.

Additionally, we have confirmed, for both X-ray and UV excitations, that the appearance of the red-shifted peak at lower temperature showed a reversible nature even after several heating and cooling cycles (see SI section 7). This indicates the lack of irreversible changes to the sample by the X-ray excitation, ruling out alternative explanations to collective emission such as defect emission[93] or superlattice deterioration due to ageing[79]. To provide additional validation of this conclusions, we rule out possible surface effects or substrate dependence by measuring the superlattices on 10 different samples on different substrates: 2 Kapton; 6 Kapton + 50 nm Au; 1 Kapton + 50 nm Au + 2 nm SiO$_2$; and 1 Kapton + 50 nm Pt (Fig. 3e, SI section 6). Despite sample variability, we observe collective radioluminescence with similar features in all cases, showing the robustness of this enhanced scintillation mechanism. Specifically, the averaged fast and slow lifetimes are $\tau_{\text{fast}} = 0.23 \pm 0.05$ ns and $\tau_{\text{slow}} = 0.90 \pm 0.13$ ns, respectively. These values represent a consistent 4-fold enhancement of the coupled emission rate relative to uncoupled emission at the same temperature, which is a 14-fold enhancement relative to emission at room temperature.

**Outlook**

The scalability and tunability of perovskite QDs make them especially adaptable for further innovation in scintillators technology[40]. Advances in synthesis and assembly techniques could reduce variability and enable more complex assemblies, such as layered heterostructures or complex superlattice arrangements[68], potentially combining the superfluorescent behavior with enhanced performance in nanophotonic scintillators [26,27,29], or leveraging plasmonic effects[28].

The collective emission observed in these materials establishes them as "quantum-optical scintillators", suggesting that concepts from quantum many-body science could contribute to scintillator technology. Strong dipole-dipole coupling and exciton-exciton interactions drive this behavior, and may be further enhanced by using nanophotonics. Future efforts exploring the quantum optical properties of these collective emission effects could extend our work to even higher X-ray energies and other high-energy particle excitations.

Looking forward, the observation of ultrafast X-ray-driven collective emission, with scintillation lifetimes as short as 230 ps, offers exciting opportunities for applications and for fundamental research. Specifically, these next-generation scintillators can improve temporal resolution in time-of-flight measurements, benefiting medical imaging techniques like positron emission tomography and high-energy physics detectors. Such a vision will require scaling the material fabrication to larger volumes, and mixing such QD scintillators with larger volume traditional scintillators in concepts such as metascintillators[94–96].

**Acknowledgements**

This research was supported by the Pazy Research Foundation. We thank Prof. Sergio Brovelli for illuminating discussions. We are gratefully for Dr. Maria Koifman-Khristosov for her help with SEM characterization. C. R.-C. is supported by a Stanford Science Fellowship.

# Superfluorescent scintillation from coupled perovskite quantum dots

# Supplementary Information

Shaul Katznelson[1†], Shai Levy[1,2†], Alexey Gorlach[1†], Nathan Regev[1†], Michael Birk[1], Chen Mechel[1], Offek Tziperman[1], Roman Schuetz[1], Rotem Strassberg[1,2], Georgy Dosovitsky[1], Charles Roques-Carmes[3], Yehonadav Bekenstein[1,2], and Ido Kaminer[1]

[1]*Solid State Institute, Technion – Israel Institute of Technology, Haifa 3200003, Israel*
[2]*Faculty of Materials Science and Engineering, Technion – Israel Institute of Technology, Haifa 3200003, Israel*
[3]*E. L. Ginzton Laboratory, Stanford University, Stanford, California 94305, US*
† *equal contributors*
kaminer@technion.ac.il

## S1: Sample fabrication

Substrate fabrication: The substrates were prepared by folding standard Kapton tape onto itself, followed by deposition of various materials using thermal evaporation. An adhesion layer of a few nanometers of either chromium or titanium was deposited first, followed by: (i) 50 nm of gold (substrate named Kapton+Au), (ii) 50 nm of gold with 2 nm of $SiO_2$ (substrate named Kapton+Au+$SiO_2$), or (iii) 50 nm of platinum (substrate named Kapton+Pt). Kapton-only substrates were also used for part of the samples.

Materials: Cesium Carbonate (99.9%, Aldrich), Lead Bromide (99.998%, Aesar), Octadecene (90%, Acros), Oleic acid (90%, Aldrich), Oleylamine (98%, Aldrich), Toluene (99.8%, Aldrich), Zinc Bromide (99.9%, Aesar).

All chemicals were used as purchased without further purification.

Synthesis of $CsPbBr_3$ superlattices and superlattice preparation: First, Cesium-oleate precursor was prepared in a 50 mL three-necked round-bottomed flask by dissolving $Cs_2CO_3$ (0.25 g) in a mixture of oleic acid (0.8 g) and octadecene (7 g) at 150 °C for 10 minutes under a $N_2$ atmosphere on a Schlenk line. The precursor solution of Pb and Br was prepared by dissolving PbBr2 (75 mg) and a varying amount of $ZnBr_2$ in a mixture of octadecene (5 mL), oleic acid (2 mL), and oleylamine (2 mL) in a 25 mL three-necked round-bottomed flask under a $N_2$ atmosphere at 120 °C for 10 min. After setting the temperature of the precursor solution to 200 °C, 0.4 mL of Cs precursor solution was injected to initiate the reaction. The reaction was quenched after a few seconds in an ice water bath. The product was centrifuged twice at 9000 and 3500 rpm and dispersed in clean toluene to obtain the monodisperse superlattices[S1,S2]. The superlattices were prepared on the desired substrate by a drop casting process of the solution in toluene. Silicon wafer was used as a substrate for SEM characterization, and carbon film on 300 mesh copper grids for TEM characterization. A 25 µL volume of the solution was cast and dried under ambient conditions for several hours (usually 3 to 5 hours) followed by vacuuming.

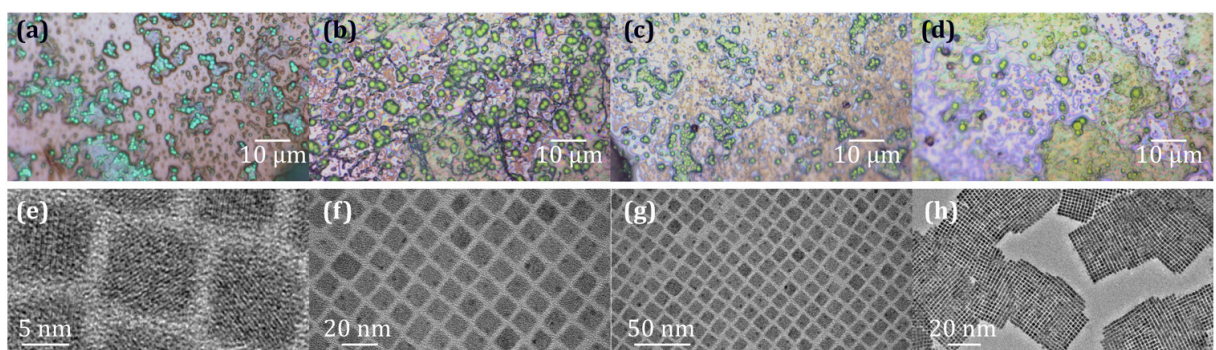

*Figure S1: CsPbBr₃ quantum dots and superlattice characterization. (a-d) Optical microscope images of CsPbBr₃ superlattices on various substrates: Kapton, Kapton+Au, Kapton+Au+SiO₂ and Kapton+Pt, respectively. (e-f) TEM images of the superlattices at various magnifications.*

## S2: Experimental setup

Radioluminescence setup: We used an IMOX X-ray tube from HELMUT FISCHER, featuring a copper target, emitting characteristic X-ray radiation at 8 keV, produced by an electron beam with adjustable settings ranging from 0-50 kV and 0-600 µA. The X-ray beam is focused by a polycapillary lens to a ~50 µm diameter spot. The X-ray radiation was directed at the sample from its back side, which was positioned inside a vertically oriented Linkam cryostage and probe station (HFS350EV-PB4) equipped with a Kapton window to minimize X-ray attenuation and evacuated to 100 mBar.

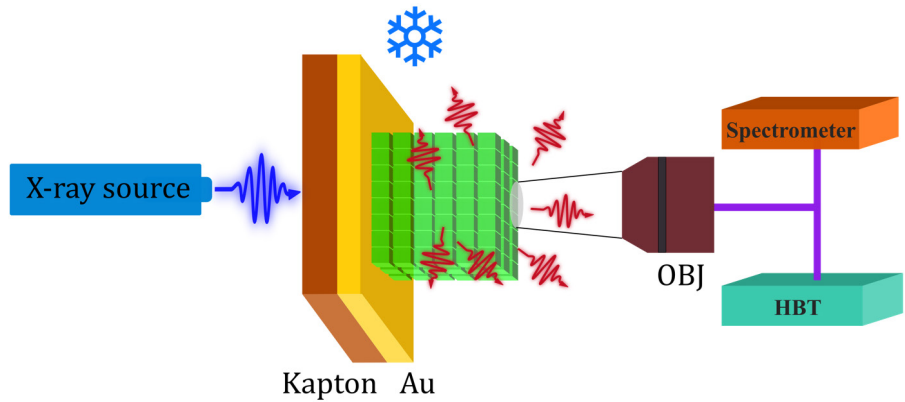

*Figure S2: Schematic of the X-ray setup* measuring radioluminescence spectrum and $g^{(2)}(\tau)$ in transmission mode.

A glass window on the back plate allows collection of the emitted light through a Motic X5 objective lens, collimating it into a 100 µm multimode fiber and directed to either a Kymera 328i spectrometer by Andor or a Hanbury-Brown-Twiss (HBT) setup. The HBT setup included an inline fiber beamsplitter, with each output connected to one of two 100 µm grade E single-photon avalanche detectors (SPADs) from MPD. The SPAD signals were time-correlated using a TimeTagger from Swabian Instruments.

UVC photoluminescence measurements were conducted with the same system, where the excitation source was changed to a KrCl excimer lamp (222nm) and the cryostat window was change to quartz to allow this wavelength to pass.

Cryogenic micro photoluminescence (PL) and time resolved photoluminescence (TRPL) measurements: Spectroscopic characterizations were performed using an Edinburgh FLS1000 spectrometer coupled to Nikon Eclipse UPRIGHT microscope with a THMS350 V Linkam temperature-controlled vacuum cryogenic stage with LNP95. All the deposited superlattices were loaded into the cryostat, evacuated to 100 mBar to prevent ice formation; during temperature changes the system had 180 s of stabilization time. The samples were excited either by an Edinburgh Xe lamp coupled with a notch filter at various wavelengths of 325 nm to 425 nm, or by an Edinburgh efficient pulse laser (EPL) of 375 nm or 405 nm coupled through an objective. The emission is collected through the same objective in reverse, using a 414 nm longpass dichroic mirror to filter out the incident laser light. TRPL measurements were performed in a multichannel scaling (MCS) mode with a monochromator used for separating the coupled and uncoupled emissions.

Transmission electron microscopy (TEM) characterization: One drop of dilute superlattices solution (1:20 dilution) was cast onto a TEM grid (carbon film only, on 300 mesh copper grid). The samples were observed in TEM mode with a Thermo Fisher/FEI Tecnai G2 T20 S-Twin LaB6 TEM operated at 200 keV, with a Gatan Rio9 CMOS camera.

Scanning electron microscopy (SEM): One drop of dilute superlattices solution was cast on a silicon substrate for SEM characterization using Zeiss Ultra-Plus FEG-SEM. Samples were measured at a working distance of 3 mm using an acceleration voltage of 5 kV.

## S3: Temperature spectral progression of photoluminescence

We observe a consistent spectral response of $CsPbBr_3$ under UV excitation across all substrates (Fig. S3). Specifically, around 220 K, a narrow redshifted peak begins to emerge and becomes more prominent at lower temperatures. This redshifted peak is attributed to superfluorescence, or collective emission from coupled quantum dots.

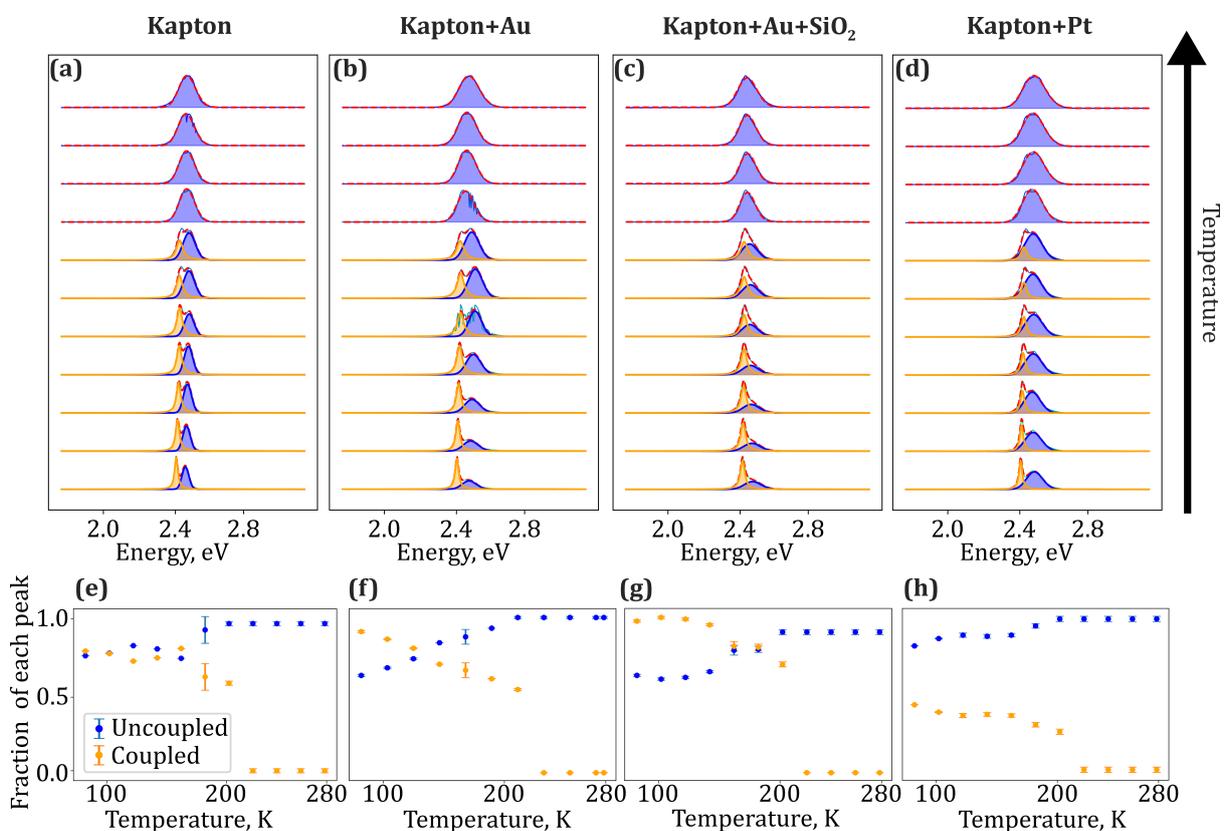

*Figure S3: Spectral response and peak weight ratio of $CsPbBr_3$ under UV excitations for various substrates. (a-d) Spectral response at different temperatures for samples on substrates: Kapton, Kapton+Au, Kapton+Au+SiO$_2$, and Kapton+Pt, respectively. (e-h) Relative weight of each peak at different temperatures for samples on substrates: Kapton, Kapton+Au, Kapton+Au+SiO$_2$, and Kapton+Pt, respectively.*

## S4: Temperature spectral progression of radioluminescence

The spectral response of $CsPbBr_3$ under X-ray excitation, showing similar features across all substrates (Fig. S4). Around 220 K, the same temperature as in Fig. S3, a broad redshifted peak begins to appear and becomes more dominant at lower temperatures. This redshifted peak is attributed to collective emission.

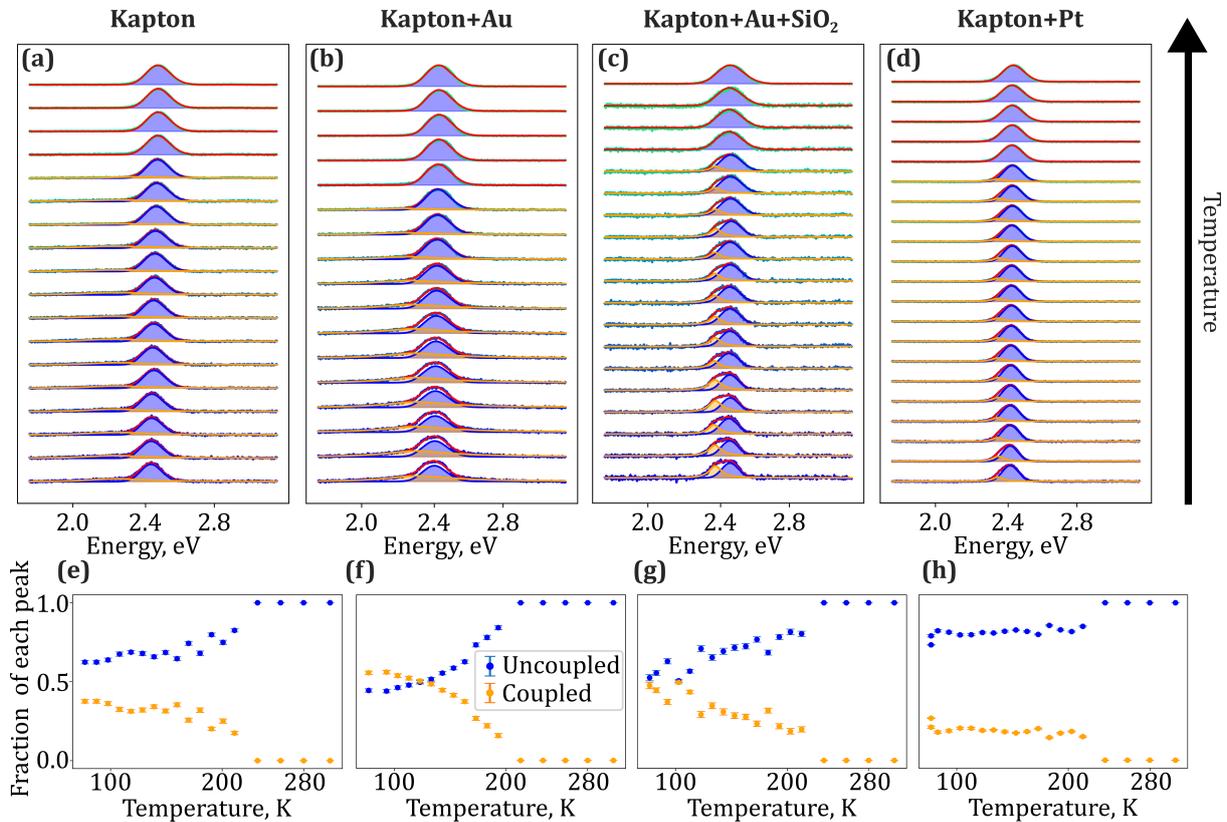

*Figure S4: Spectral response and peak weight ratio of CsPbBr3 under X-ray excitations for various substrates. (a-d) Spectral response at different temperatures for samples on substrates: Kapton, Kapton+Au, Kapton+Au+SiO₂, and Kapton+Pt, respectively. (e-h) Relative weights of each peak at different temperatures for samples on substrates: Kapton, Kapton+Au, Kapton+Au+SiO₂, and Kapton+Pt, respectively.*

## S5: Spectral response of radioluminescence for different collection areas

In this set of measurements, a mechanical iris was placed in the beam path after the collection objective and before coupling to the optical fiber. By adjusting the iris diameter, the effective collection area of the sample was modified. Despite this change, Fig. S5 shows that the widths of the spectral peaks remain largely unchanged under different collection areas, with the largest collection area shown in Fig. S5a and the smallest in Fig. S5c . This test rules out substantial variation in QD properties across the samples, which would have shown narrower spectral peaks from smaller collection areas in Fig. S5e.

Interestingly, smaller collection areas yield slightly higher fractions of emission from coupled QDs (Fig. S5d), likely due to the exclusion of rogue QDs outside the main ensemble of superlattices, which contribute only to spontaneous emission and not to collective emission.

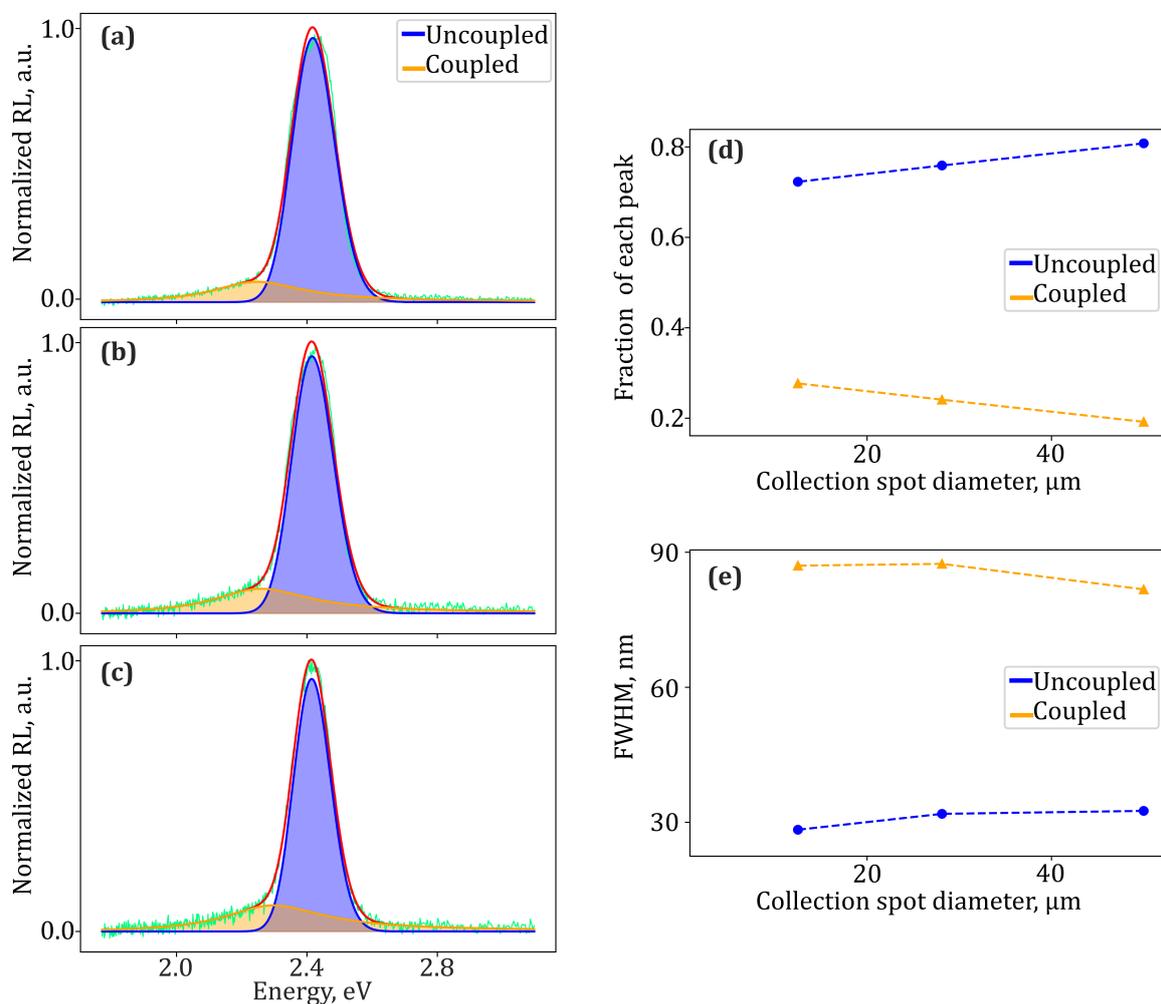

*Figure S5: Effects of collection area on the ratio between spontaneous and collective emission. (a-c) Spectrum of X-ray-driven $CsPbBr_3$ on a Kapton substrate at 80 K, measured for decreasing collection spot diameters, 50, 28, and 13 μm, respectively. (d) Extracted relative peak weights as a function of collection spot diameter. (e) Extracted full width at half maximum (FWHM) of each spectral peak as a function of collection spot diameter.*

## S6: Radioluminescence $g^{(2)}(\tau)$ and extracted lifetimes for different samples at 80K

We present the HBT analysis results for different samples under X-ray irradiation (Fig. S6). The lifetimes are extracted by fitting double-exponential curves. The fits are constrained to maintain the same relative peak weights as observed in the corresponding spectra (shown in the insets) for each sample. These results are summarized in Figure 3c of the main text.

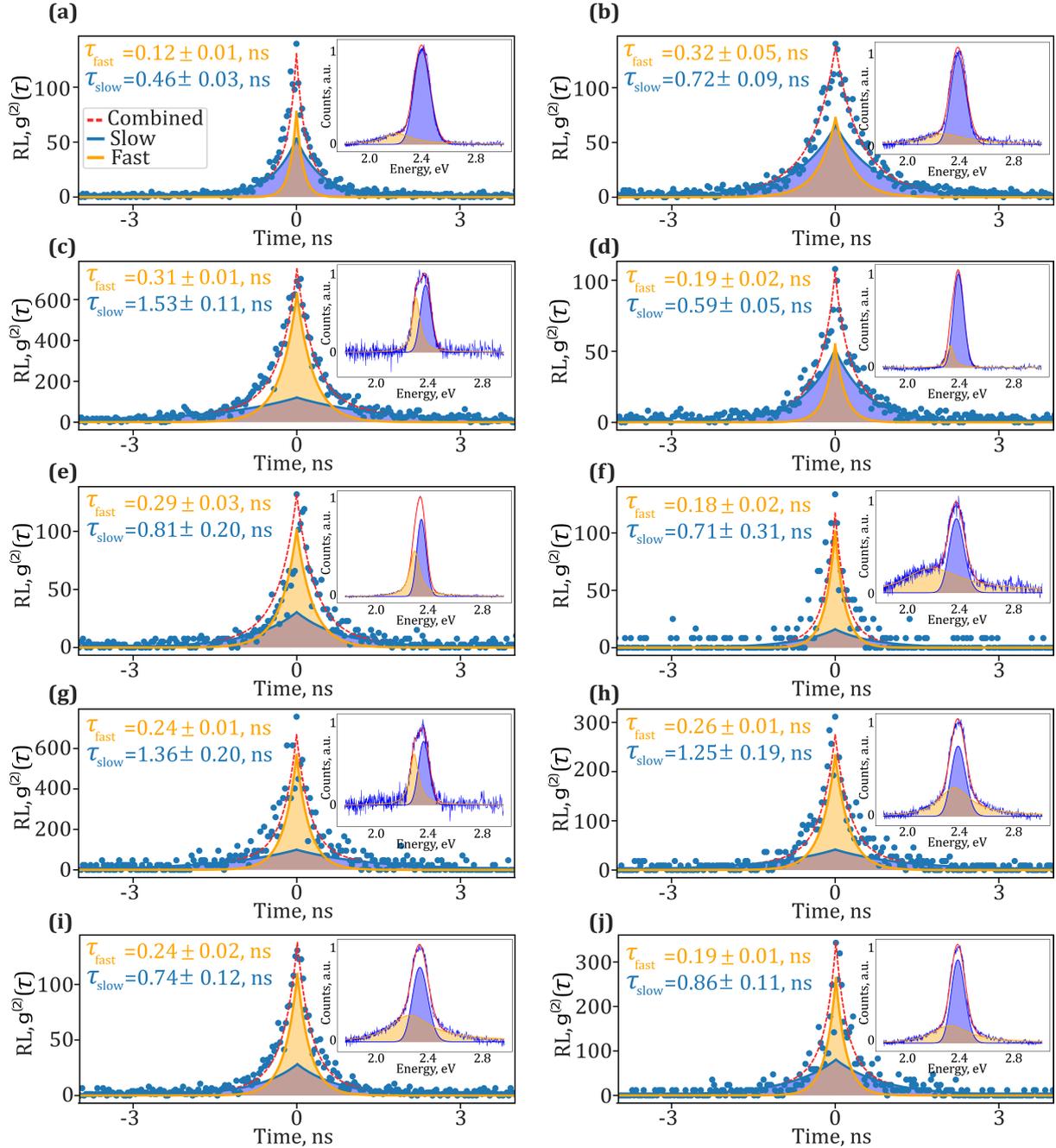

*Figure S6: $g^{(2)}(\tau)$ and extracted lifetimes for X-ray-driven CsPbBr₃ superlattice on different substrates at 80 K. (a-b) Kapton substrate. (c) Kapton+Au+SiO₂ substrate. (d) Kapton+Pt substrate. (e-j) Kapton+Au substrate. The insets show the corresponding spectra from which the relative weights are extracted.*

## S7: Robustness of the collection emission phenomenon under cooling cycles

The samples preserve their optical properties, including their low-temperature collective emission from coupled QDs, when kept in inert states (vacuum desiccator), even withstanding several cooling-heating cycles, as shown in Fig. S7. The figure displays three measurements of the same sample, showing similar results: The initial measurements (Fig. S7a1,2), measurements 24 days later (Fig. S7b1,2), and measurements a day after that (Fig. S7c1,2). The sample was kept in room temperature between measurements.

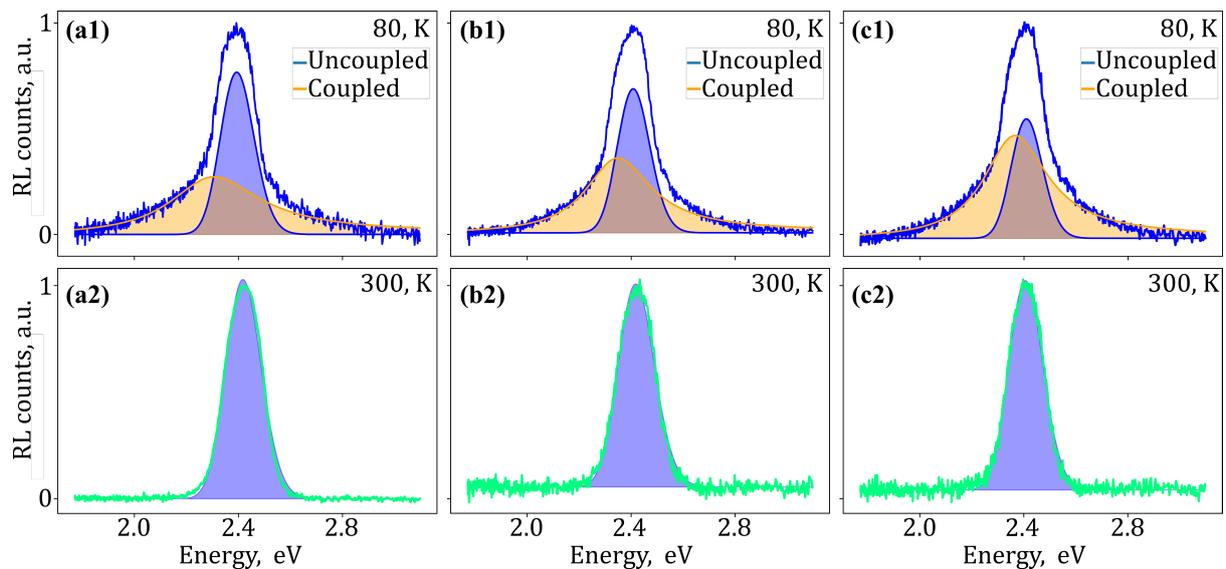

*Figure S7: Repeatability of collective emission from coupled QDs after several cooling cycles. (a1,2) Emission spectra of X-ray-driven $CsPbBr_3$ superlattices at 80 K and 300 K, respectively. (b1,2) Emission spectra of the same sample in another cooling cycle, measured 24 days later. (c1,2) Emission spectra of the same sample on the following day, still showing collective emission at 80 K.*

## S8: Theory of collective emission driven by X-ray and UV excitations

In this section, we present a qualitative model of collective emission in perovskite QDs, capturing all the measured features for excitation by both UV and X-ray radiation. The mechanism of collective emission is the same in both cases: Strong dipole-dipole interactions between QDs couple them, making them behave in the collective manner, inducing redshift in the emission spectrum and accelerating the emission rate. This type of collective behavior of QDs is known in the literature as superfluorescence and its theoretical description matches experiments in molecular aggregates[S3], including J- and H-aggregates[S4].

Here we develop an approach based on such conventional theories of superfluorescence[S3,S4]. We extend the theory to capture the case of X-ray excitations by accounting for multiple simultaneous incoherent excitations in close proximity to each other (in comparison, each UV photon can only produce one excitation). We capture this regime by considering the simultaneous excitations as increasing the dipole-dipole interaction strength and its variability, such that each individual excitation is already modified. We can then use the effective single-excitation theory below to account for both the UV and X-ray regimes. This generalization of the basic theory of superfluorescence enables describing measurements of collective emission across several orders of magnitude of excitation energies, from the optical to the X-ray regime.

We consider an initially localized single excitation, since coherence is lost in the excitation process. This single localized excitation can be at any of $N$ QDs, not knowing which of the QDs is exactly excited:

$$\rho(0) = \frac{1}{N}\sum_{n=1}^{N}|n\rangle\langle n|, \tag{S1}$$

where $|n\rangle$ is

$$\begin{cases} |1\rangle = |\text{egggg}\ldots\text{g}\rangle, \\ |2\rangle = |\text{gegg}\ldots\text{g}\rangle, \\ \quad\ldots \\ |N\rangle = |\text{gggg}\ldots\text{e}\rangle. \end{cases} \tag{S2}$$

This initial state is completely incoherent, having a minimal overlap with superradiant Dicke states. Thus, the collective optical effects emerge purely from QDs interactions. Such a process is usually attributed to superfluorescence[S3,S4], separating it from other forms of superradiance.

The dynamics of the incoherent state Eq. (S1) is described by a Lindblad master equation considering dipole-dipole interactions between the QDs as well as the emission process:

$$\frac{d\rho}{dt} = -i[H,\rho] + \sum_{ij}\Gamma_{ij}\left[\sigma_i^-\rho\sigma_j^+ - \frac{1}{2}\{\sigma_i^+\sigma_j^-,\rho\}\right], \tag{S3}$$

with $H = \sum_i \omega_i \sigma_i^+ \sigma_i^- + \sum_{i\neq j} J_{ij} \sigma_i^+ \sigma_j^-$, where $\sigma_i^\pm$ are the Pauli matrices of the $i$th QD. Here $\omega_i$ is the frequency of $i$th QD, with inherent variability described by a Gaussian probability distribution $P(\omega) = (2\pi\sigma_E^2)^{-1/2}\exp(-(\omega-\omega_0)^2/2\sigma_E^2)$, where $\omega_0$ is the central frequency and $\sigma_E$ is the standard deviation. $\Gamma_{ij}$ is the emission matrix that characterizes collective emission. Assuming the QDs are closely packed in superlattices, we can set $\Gamma_{ij} = \Gamma_0$, where $\Gamma_0$ is the spontaneous emission rate. The interaction between the $i$-th and $j$-th QDs is described by $J_{ij}$. We find that considering just the nearest neighbor interactions ($J_{ij} \neq 0$ only for neighbor QDs) is sufficient to capture all the main features of the experiments under both UV and X-rays and for the entire range of temperatures.

We are interested in finding the spectrum and time-dependent intensity of the emission, since these are the main observables in our experiments. These properties can be derived from the first-order coherence, found using Quantum Regression Theorem[S5]:

$$G^{(1)}(\tau+t,t) = \sum_{ij}\Gamma_{ij}\text{Tr}\left[\sigma_+^j\left(\sigma_-^i\rho(t)\right)(\tau)\right]. \tag{S4}$$

Evaluating $G^{(1)}(\tau+t,t)$ requires first evolving the matrix $\rho(0)$ for time $t$ using Eq. (S3) and then evolving the matrix $\sigma_-^i\rho(t)$ for an additional time $\tau$. The spectrum $S(\omega)$ and time-dependent intensity $I(t)$ of the emitted light can then be found as:

$$I(t) = G^{(1)}(t,t), \qquad S(\omega) = \int_0^\infty dt \int_0^\infty e^{-i\omega\tau} d\tau \; G^{(1)}(\tau+t, t). \tag{S5}$$

Substituting the above equations, we can express $I(t)$ and $S(\omega)$ in terms of $\Gamma_{ij}$ and $H_{ij}$, expressed in matrix forms as $\hat{\Gamma}$ and $\hat{H}$:

$$\begin{cases} I(t) = \text{Tr}\left[\hat{\Gamma} e^{\left(-\frac{\hat{\Gamma}}{2} - i\hat{H}\right)t} e^{\left(-\frac{\hat{\Gamma}}{2} + i\hat{H}\right)t}\right], \\ S(\omega) = \text{Re}\left\{\int_0^\infty dt \int_0^\infty e^{-i\omega\tau} d\tau \; \text{Tr}\left[\hat{\Gamma} e^{\left(-\frac{\hat{\Gamma}}{2} - i\hat{H}\right)t} e^{\left(-\frac{\hat{\Gamma}}{2} + i\hat{H}\right)t} e^{\left(-\frac{\hat{\Gamma}}{2} + i\hat{H}\right)\tau}\right]\right\}. \end{cases} \tag{S6}$$

Differences between collective emission under UV and X-ray excitations

The theory above describes both spontaneous and collective emission under both UV and X-ray excitations. In room temperature, the interactions $J_{ij}$ are weak, and the dynamics is dominated by the variability in frequency, $\sigma_E$, also known as inhomogeneous spectral broadening, which suppresses coherent effects. We can thus estimate the parameters $\hbar\omega_0 = 2.45$ eV and $\sigma_E = 100$ meV from the measured emission from uncoupled (individual) QDs in the room-temperature spectrum, for both X-ray and UV excitation (violet curve in Fig. S8 a,b).

Dipole-dipole interactions are essential for generating collective effects, for both UV and X-ray excitations. The interactions synchronize the QDs, leading to superfluorescence (Fig. S8 c-f red curve) with faster lifetime and a red-shifted spectral peak. However, the dipole-dipole interactions differ significantly between UV and X-ray excitations. For UV excitation, we assume no variation in dipole-dipole interaction strengths since each UV photon produces one excitation (Fig. S8b, blue curve), which then interacts uniformly with the neighboring (unexcited) QDs. An average interaction strength $J_0 = 50$ meV and standard deviation $\sigma_J = 0$ meV fit typical measurements.

In contrast, an incident X-ray photon produces multiple simultaneous excitations of different QDs with random dipole orientations within a small volume that includes strong interactions. We capture this effect by an effective theory that considers a single collective excitation with a larger average dipole-dipole interaction strength $J_0 = 150$ meV and a larger variation in these strengths $\sigma_J = 100$ meV, compared to the parameters for a UV excitation (Fig. S8a, blue curve). These parameters also explain the shorter lifetime under X-ray excitations, as observed in the experiment. The broader peak and faster emission for X-ray excitation (Fig. S8 c,d compared to Fig. S8 e,f) may seem counterintuitive, but it has a simple explanation: dipole-dipole coupling of any strength leads to lifetime shortening, never prolongation. Thus, averaging over different dipole-dipole couplings results in overall lifetime shortening.

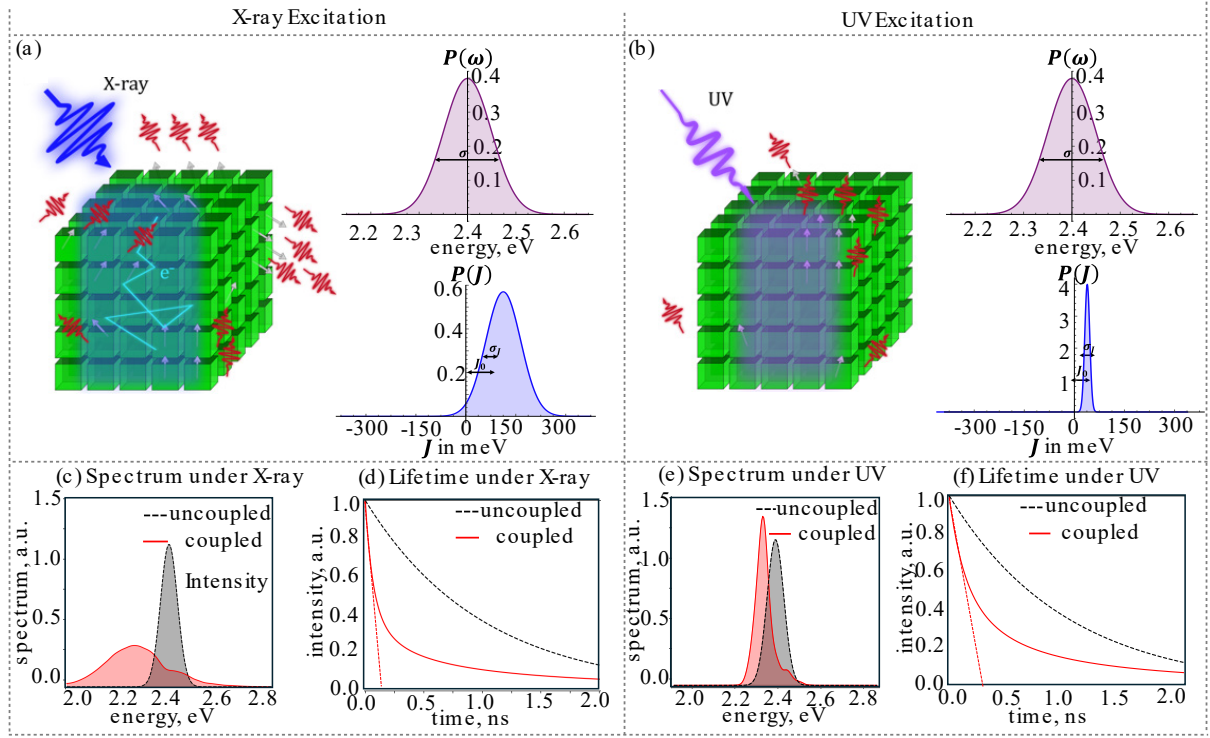

*Figure S8: Theory of collective emission and comparison of UV and X-ray excitations. (a1-3) Each X-ray photon generates multiple simultaneous excitaitons within a small volume, creating collective emission effectively captured by strong dipole-dipole interactions with large variability ($J_0 = 150$ meV, $\sigma_J = 100$ meV). (b1-3) Each UV photon generates a single excitation, creating collective emission by strong dipole-dipole interactions with no variability ($J_0 = 50$ meV, $\sigma_J = 0$ meV). (a2,b2) Emission spectra for coupled and uncoupled QDs for X-ray and UV excitations, respectively. (a3,b3) Emission intensity over time for coupled and uncoupled emission for X-ray and UV excitation, respectively.*

Our theory also applies to other features of collective coherence measured in other works, such as the $G^{(1)}$ coherence observed in[6]. In our case, due to the continuous wave X-ray excitation, the $G^{(1)}$ coherence is related to the spectrum that we measured. Since the emission by coupled QDs for X-ray excitation has a broader spectral peak than that emitted from uncoupled QDs, the $G^{(1)}$ measure of emission by coupled QDs will be shorter than that by uncoupled QDs. This result is opposite to the case of UV-driven emission from coupled QDs[S6]; thus, $G^{(1)}$ is not as useful to characterize collective emission by coupled QDs under X-ray excitation. It remains to be seen whether other mechanisms of collective emission reported in the literature can also be captured by our quantum optical framework or require its generalization.

### S9: Monte-Carlo simulations

To capture the process of X-ray photon absorption and production of excitations, we use the Geant4[S7–S9] Monte-Carlo simulation. The simulation assumes an X-ray photon incident on the sample, generating a photoelectron that produces localized excitations of QDs along its path. The simulation records all the major events of the emission process as exemplified in Fig. S9b. The statistical analysis is shown in Figs. S9ce, highlighting that neighboring sites are excited with higher likelihood than distant sites. This spatial correlation of excitation supports our model of interactions among multiple simultaneous incoherent excitations, which explains the measurements of collective radioluminescence. The correlated neighboring excitations analyzed here are then used in the Lindblad master equation formalism to explain the differences between the X-ray-driven and UV-driven collective emission.

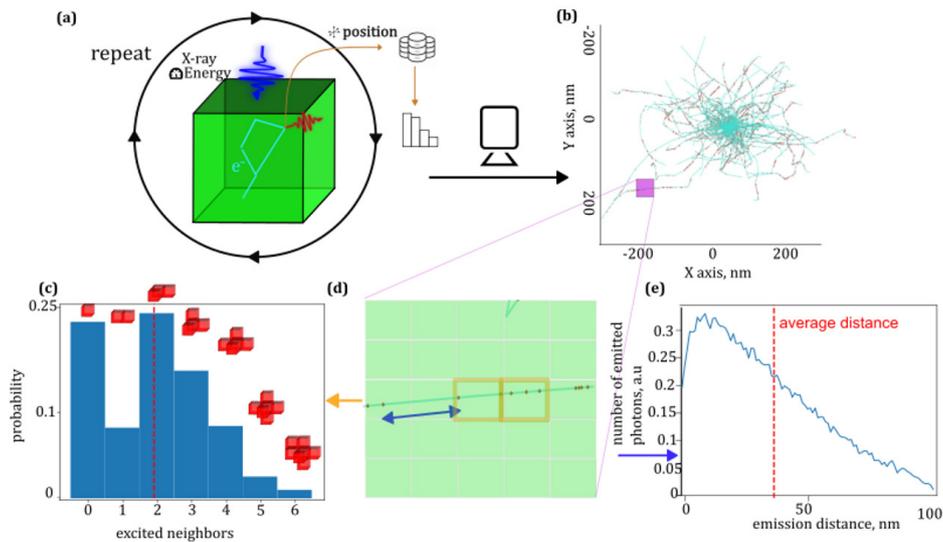

*Figure S9: Monte-Carlo simulation process. (a) Schematics of the simulation, marking the incident X-ray photon, generated photoelectron, and optical photon emission. (b) Example simulation results for an incident 8 keV X-ray photon. The cyan lines represent the path of the photoelectrons excited by the X-rays, and the red dots represent excitation sites of QDs. The X-ray photons enter the material in the normal direction to the plane of the drawing. (c) Probability of neighbor excitation. The histogram shows the probability of any number of neighboring QDs of an excited QD to also be excited. (d) Zoom-in view of emission sites. The green rectangles provide a scale by representing the superlattice, each cell being 10 nm, representing the periodicity between QDs. Blue double-sided arrow denotes a distance between two localized excitations, while orange rectangles mark two neighboring QD sites. (e) Distribution of distance between excitations. The x axis is the mean distance between different excitations produced by the same incident X-ray photon.*

The simulation of Fig. S9 uses the following parameters: uniform distribution of $CsPbBr_3$ (respective Z-numbers of 55,82,35). The geometry consists of a bulk of $CsPbBr_3$ with a density of 4.85 $\frac{g}{cm^3}$. In each iteration of the simulation, a single X-ray photon with an energy of 8 keV impinges the sample, producing hundreds of localized excitations. The simulation was repeated 10,000 times and the concatenated data was analyzed in Figs. S9c,e.

This simulation does not capture the propagation of the optical photons, instead finding them as localized excitation points that are no longer treated by the simulation, instead providing the distribution for the next simulation step.

Additional filters that were placed on the simulation are low energy filters, removing from the illustration (not from the analysis) electrons with energy of less than 100 eV. We also applied a filter removing photons that were generated directly from the X-ray absorption, a process which is possible in principle (e.g. Compton scattering) but not likely for our physical parameters.